# Evolution of primordial planets in relation to the cosmological origin of life

N. Chandra Wickramasinghe[a]*, Jamie H. Wallis[1], Carl H. Gibson[2] and Rudolph E. Schild[3]

[1]Cardiff Centre for Astrobiology, 24 Llwynypia Road, Cardiff, CF14 0SY, UK
[2]University of California San Diego, La Jolla, CA 92093-0411, USA
[3]Center for Astrophysics, 60 Garden Street, Cambridge MA 02138, USA

*Corresponding author: Email: ncwick@gmail.com, Wickramasinghe@cf.ac.uk

## ABSTRACT

We explore the conditions prevailing in primordial planets in the framework of the HGD cosmologies as discussed by Gibson and Schild. The initial stages of condensation of planet-mass H-$^4$He gas clouds in trillion-planet clumps is set at 300,000 yr (0.3My) following the onset of plasma instabilities when ambient temperatures were >1000K. Eventual collapse of the planet-cloud into a solid structure takes place against the background of an expanding universe with declining ambient temperatures. Stars form from planet mergers within the clumps and die by supernovae on overeating of planets. For planets produced by stars, isothermal free fall collapse occurs initially via quasi equilibrium polytropes until opacity sets in due to molecule and dust formation. The contracting cooling cloud is a venue for molecule formation and the sequential condensation of solid particles, starting from mineral grains at high temperatures to ice particles at lower temperatures, water-ice becomes thermodynamically stable between 7 and 15 My after the initial onset of collapse, and contraction to form a solid icy core begins shortly thereafter. The icy planet core, which includes a fraction of radioactive nuclides, $^{26}$Al and $^{60}$Fe, melts through interior heating. We show, using heat conduction calculations, that the interior domains remain liquid for tens of My for 300km and 1000km objects, but not for 30 or 50km objects. Primordial-clump-planets are separated by ~ 1000 AU, reflecting the high density of the universe at 30,000 yr. Exchanges of materials, organic molecules and evolving templates readily occur, providing optimal conditions for an initial origin of life in hot primordial gas planet water cores when adequately fertilized by stardust. The condensation of solid molecular hydrogen as an extended outer crust takes place much later in the collapse history of the protoplanet. When the object has shrunk to several times the radius of Jupiter, the hydrogen partial pressure exceeds the saturation vapour pressure of solid hydrogen at the ambient temperature and condensation occurs.

## 1. INTRODUCTION

The origin of life in the form of self-replicating cells capable of further evolution requires organic molecules to be present within liquid environments that remain stable for tens of millions of years. These are necessary, but not sufficient conditions for the emergence of life. Even with the provision of cosmic niches that meet the necessary conditions, the probability of the emergence of life from non living chemicals is, in our view, so minuscule as to demand the biggest available cosmological setting (Hoyle and Wickramasinghe, 1982; Joseph, 2001). In the HGD cosmological model developed by Gibson (1996) and Schild (1996) the best venues for this process to operate are primordial planets that begin forming as initial density fluctuations nearly 300,000 yr after the Big Bang (Gibson, Schild and Wickramasinghe, 2010).

When the cosmological plasma begins to recombine to form a neutral gas, comprised mainly of H and He, instabilities develop over two principal scales – one corresponding to masses of ~ $10^{29}$g and another corresponding to masses ~$10^{39}$g. The former is appropriate to planetary masses and the latter to masses of globular clusters. In their model, Gibson and Schild suggest that the earliest stars form by run-away accretion/coagulation of planetary mass clouds. In this paper we concentrate on the smaller mass gas clouds that form initially and show that they provide the best niches within a cosmological setting for prebiotic chemistry and life to develop.

## 2. AMBIENT DENSITY AND TEMPERATURE

In the approximation of a Euclidian universe the cosmological scale factor (size of the Universe) R(t) at epoch t after the Big Bang is proportional to $t^{2/3}$, thus giving a density proportional to $t^{-2}$. Therefore we have for a density-time relation:

$$\frac{\rho(t)}{\rho(t_0)} = \left(\frac{t}{t_0}\right)^{-2} \tag{1}$$

where ρ(t) represents the density and $t_0$ is the present epoch, $t_0$=13.7Gy. With the present day density taken to be $\rho(13.7Gy) \approx 10^{-30}$ g cm$^{-3}$ we obtain

$$\rho(t) = 10^{-30}\left(\frac{t}{13.7 \times 10^9}\right)^{-2} \text{ g cm}^{-3} \tag{2}$$

for any general value of time t. Setting $t=3\times 10^5$ yr we obtain $\rho(300,000 \text{ yr}) \approx 2\times 10^{-21}$ gm cm$^{-3}$. The variation of ρ with density according to equation (2) is plotted in Fig.1.

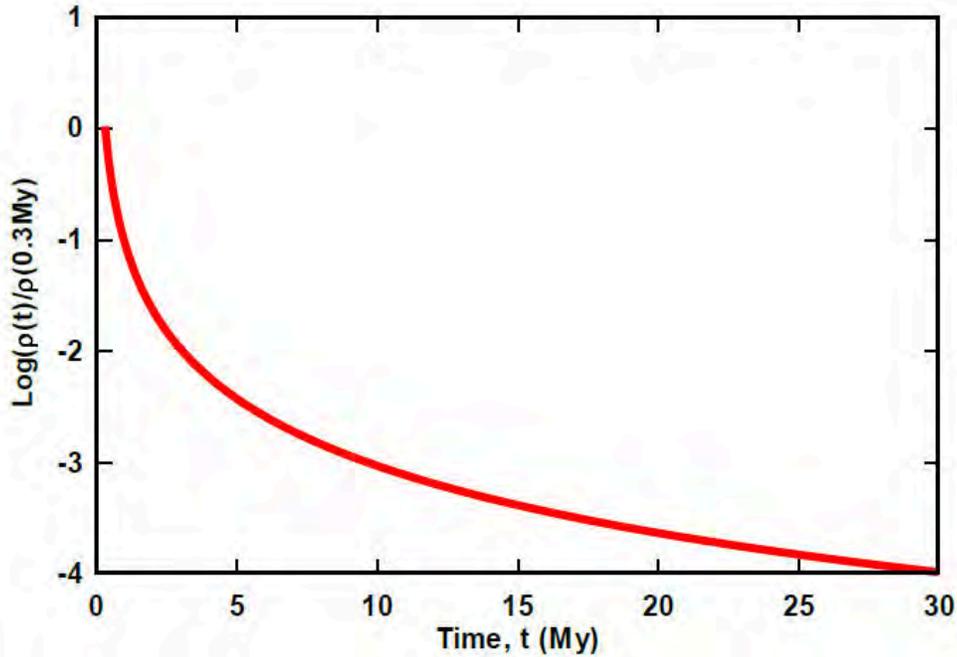

Fig. 1 Density of the universe relative to that at t = 0.3My

For notional density fluctuations of say ~ 50%, at the stage when instabilities set in, a planet-mass fragment would correspond to a sphere of radius ~300AU ($4.5\times 10^{15}$cm) and its free-fall gravitational collapse time will be

$$t_{coll} \approx \left(\frac{3\pi}{32G\rho}\right)^{\frac{1}{2}} \cong 1.1 \times 10^6 \, yr$$

The ambient gas temperature in the Universe after matter and radiation become decoupled is related to density by the relation

$$\rho \propto T^3 \tag{3}$$

giving

$$\frac{\rho}{10^{-30}} \cong \left(\frac{T}{2.7}\right)^3 \tag{4}$$

Combining (4) with (2) we get

$$T = 1.14 \times 10^3 \left(\frac{t}{10^6 \, yr}\right)^{-2/3} \tag{5}$$

so that the gas temperature at the early stages of collapse towards a frozen planet at t ≈ $10^6$ yr will be T ~ 1000K.

### 3. COLLAPSING PLANETARY CLOUD

Following the onset of plasma instabilities in the HGD models of Gibson and Schild, a gas cloud of planetary mass (an incipient frozen planet) will begin to decrease in radius. Initially, the temperature within the cloud would equal that of the exterior cosmological medium, the cloud itself remaining at relatively low densities corresponding to a few thousand to hundreds of millions of atoms per cubic centimetre. Initial collapse of the cloud can be supposed to proceed through a series of polytropic configurations. To a first order of approximation it could also be assumed that the collapse would occur in a manner that maintains an isothermal condition, with an approximate equality of temperature with the exterior medium. This situation will be expected to continue even with a diminution of radius by a factor of 1000 – ie with the cloud decreasing its radius from 300AU to 0.3AU. At some point after this internal opacity due to molecule formation plays a decisive role in weakening the thermal coupling to the exterior gas, and temperature and pressure will begin to rise towards the centre of the cloud.

With run-away accretion into larger mass stars taking place simultaneously in a fraction of these protoplanetary clouds, stellar/supernova nucleosynthesis would be expected to occur (Schild and Gibson, 2008). This would lead to the enrichment of neighbouring incipient protoplanets with a compliment of heavy elements. In addition to H and He the primordial proto planets would contain heavy elements – C, N, O, Si, P…- that could eventually condense into solid particles, and also be involved in the onset of prebiotic chemistry. This enrichment process would take place against the background of an expanding universe in which the ambient temperature continues to fall below 1000K in accordance with equation (5). Fig 2 shows the temperature of the cosmological medium as a function of time.

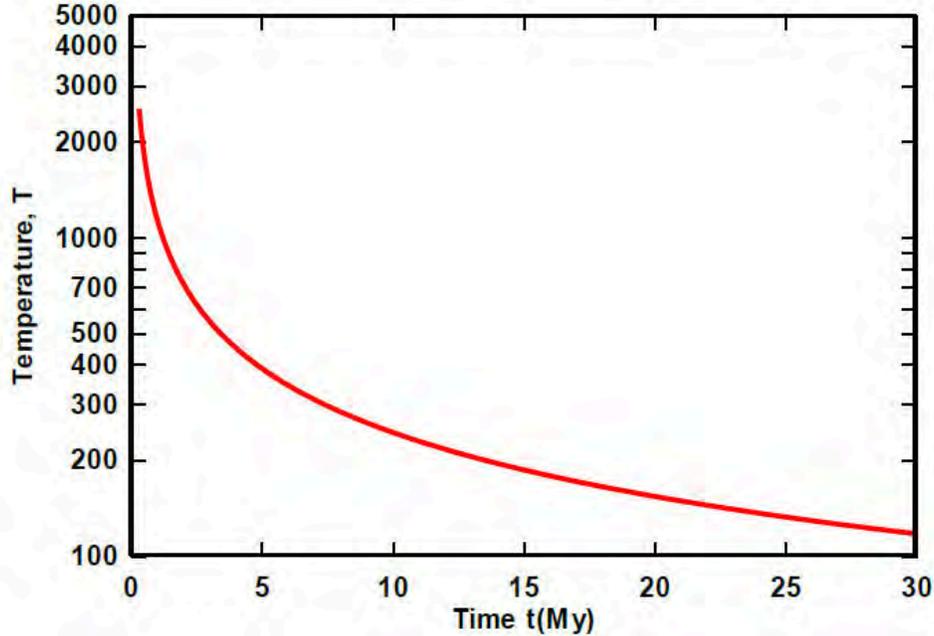

Fig.2. The ambient temperature of the universe in degrees Kelvin as a function of epoch for t ≥ 0.3My

In the early phases of isothermal collapse Fig. 2 also represents the temperature within the contracting cloud to a good approximation. Once supernovae in the vicinity have significantly enriched the gas of the protoplanet with heavy elements, including radioactive nuclides, Fig 2 can be taken to define a condensation sequence in the cooling gas cloud, with a starting temperature of ~ 1000K, density $\rho \approx 3\times10^{-21}$ gm cm$^{-3}$ and radius ~ $4.8\times10^{15}$cm (~300AU).

## 4. CONDENSATION OF SOLID PARTICLES

When molecules begin to form, the thermochemistry of the collapsing cloud would not be unlike that relevant for models of planetesimal formation in a collapsing solar nebula. The earliest condensates to form are mineral grains, metallic oxides and silicates, their partial pressures in the gas exceeding the relevant saturation vapour pressures at temperatures < 1000K. Such refractory particles would form less than one percent of the largely ice material that condenses subsequently at lower temperatures.

Note that according to the HGD cosmology, only a small fraction of planets are formed in a typical solar nebula ($3\times10^{-5}$ %). Most planets of the galaxy were formed primordially from hot (3000 K) H-He gas clouds in PGC clumps, each comprised of a trillion Earth-mass gas cloud fragments. This was the case before the first stars appear in the Universe, and the clouds remain in pristine state for the time it takes for stars to form within the PGCs, and for the most agitated PGCs (those near the protogalaxy-core centres) to "overeat" gas cloud planets and evolve into massive stars that explode as supernovae. The chemical elements C, N, O, P, Si, Ca, Fe, Ni were formed in stars and their supernovae within PGCs were then dispersed among, and gravitationally collected by, the trillion planet clouds. In this way the planet clouds were "fertilized" while they were hot and most capable of creating life chemicals and life itself. The last step in the logic requires water molecules to form and cool below the water critical temperature (374 °C, 647 K) and for organic molecules to dissolve in an aqueous medium. Thus, we present the following solar nebula calculation as an important exercise in planetary chemistry with wider cosmological applications that need to be pursued. Analysis of corresponding sophistication has not been carried out for primordial gas planets as they collapse.

A typical solar nebula calculation for a cooling gas cloud is shown in Fig. 3 (Hoyle and Wickramasinghe, 1969). The dashed segments correspond to condensation of solid particles with a consequent decrease of relevant gas phase densities.

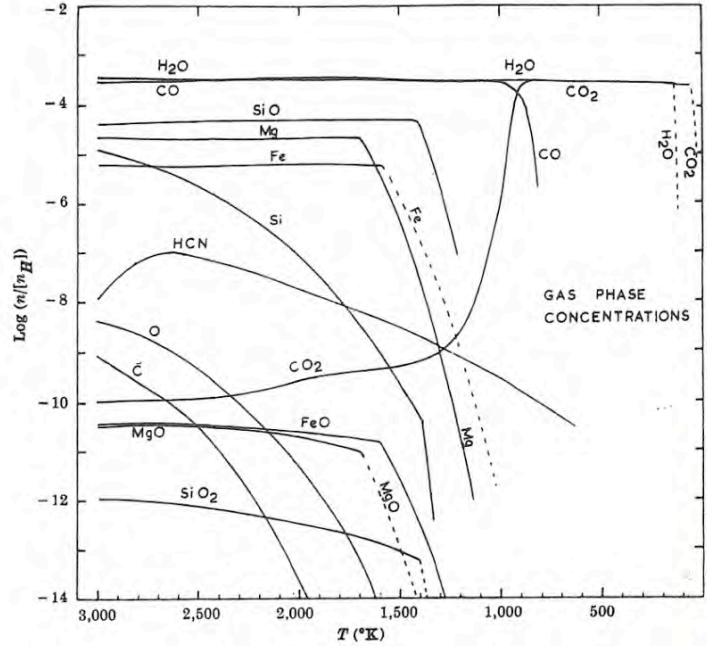

Fig 3  Gas phase equilibrium concentrations in a cooling solar nebula model

With ~ 1% of the hydrogen mass density in the form of $H_2O$, and with a O/C ratio similar to that in the solar system, chemical equilibrium calculations show that the bulk of O will be in the form of $H_2O$ for temperatures T < 500K (Hoyle and Wickramasinghe, 1969). This happens according to equation (5) for epochs t > 3.44My. However, condensation of $H_2O$ gas into solid ice requires supersaturation of the vapour – that is, the vapour pressure of $H_2O$ must exceed the saturation vapour pressure of the solid.

The saturation vapour pressure $p_{sat}$ of $H_2O$ ice is given to a good approximation by

$$\log p_{sat} \cong -\frac{2480}{T} + 4\log T + 4.06 \tag{6}$$

where T is the temperature (van de Hulst, 1946). The condensation of ice particles would thus begin when the supersaturation ratio satisfies the condition

$$\frac{10^{-2}\rho kT/m_H}{p_{sat}} > 1 \tag{7}$$

where k is Boltzman's constant, $m_H$ is the mass of the hydrogen atom and $\rho$ is the total gas density in the collapsing cloud, the factor $10^{-2}$ taking account of the mass fraction of water.

We can now estimate condensation temperatures for water-ice particles in a protoplanetary cloud embedded in the expanding universe. The ambient density of the universe is given by equation (2) (see also Fig.1), but a collapsing protoplanetary cloud containing $H_2O$ gas would have its density compressed and enhanced to varying degrees as isothermal collapse proceeds. Strictly the condition of uniform density and temperature would need to be modified using appropriate polytrope models for the collapsing cloud, with temperature and pressure rising toward the centre. However the outer optically thin layers of the cloud would continue to maintain an isothermal condition and a balance of temperature with the external background, and it is in these layers that we shall suppose condensations would occur.

The curves in Fig 4 show the supersaturation ratio calculated according to equation (6) and (7) and using equation (5) to connect T and the epoch t. The dashed curve corresponds to the reference condition of marginal supersaturation. Condensations would occur when the supersaturation ratio crosses this line from below. The red curve corresponds to a radial compression of the collapsing cloud by a factor $10^6$, the radius of the cloud being $3 \times 10^{-4}$ AU; the bottom curve is for a cloud of radius 3AU. From Fig. 4 we see that condensation of ice particles occurs when $t \approx$ 7-15 My after the Big Bang, for compressions in the range f = 10,000 to 1,000,000. The required enhancements of density are not unreasonable to be achieved towards the centre of a collapsing spherical isothermal gas cloud within the framework of a classical polytrope model with poytropic index n=∞ (Eddington, 1928).

We shall discuss this question further in Section 7. For the present purposes we shall take the ice condensation condition to be achieved at any epoch in the range 7-15My.

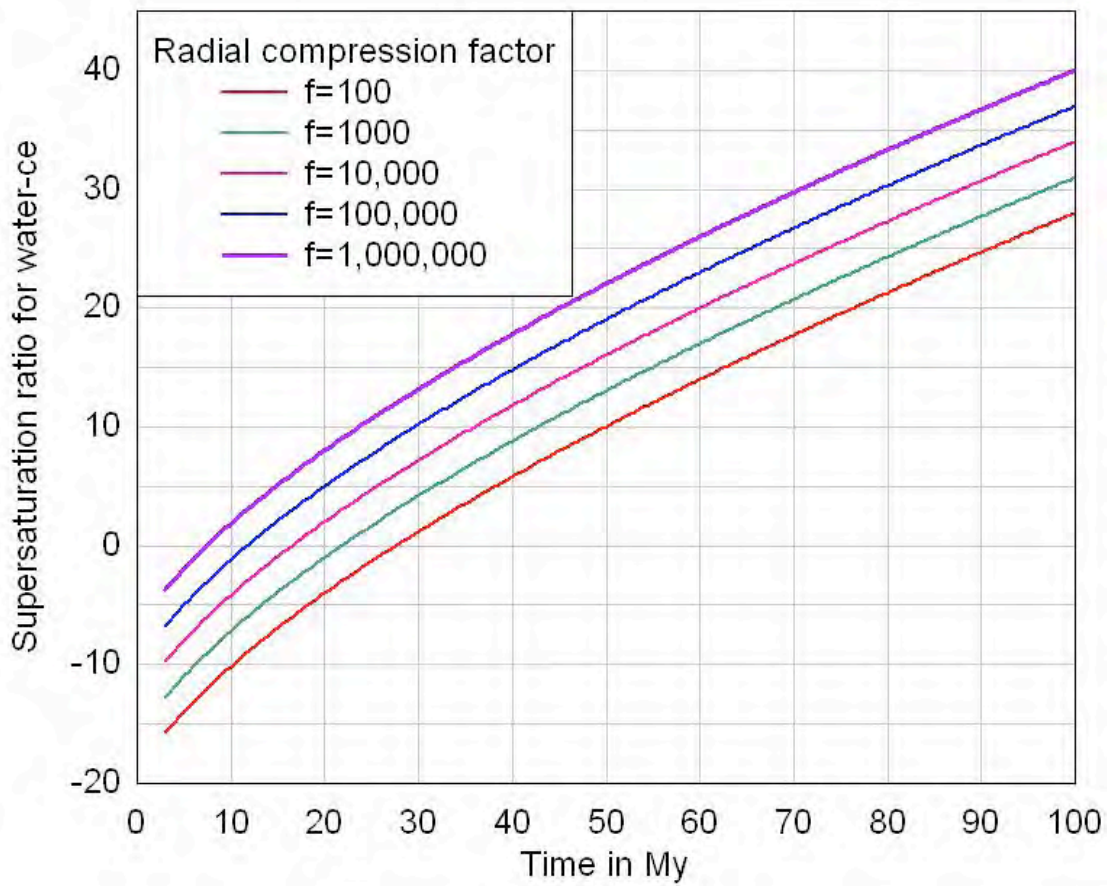

Fig. 4 The dashed black curve corresponds to unit supersaturation of ice. The several curves are for radial compressions by a factor $10^2$ to $10^6$. The oxygen in the cloud is assumed to be in the form of $H_2O$.

## 4.1 Growth of ice grains

Ice condensation will take place on pre-existing condensation nuclei, most likely on silicate or mineral grains that had formed earlier. The rate of growth of an ice grain (assumed spherical) is given by

$$\frac{dr}{dt} = \alpha \frac{p}{s} \left(\frac{m}{2\pi kT}\right)^{1/2} \tag{8}$$

Here $p$ is the partial pressure of $H_2O$, $s$ is the bulk density of ice ~ 1 g cm$^{-3}$, $m$ is the mass of an $H_2O$ molecule, $T$ is the temperature and $\alpha$ is a sticking coefficient assumed unity for the present purpose.

For a cloud that has contracted to a radius R from its initial radius $R_0 = 300$AU, we have a density $\rho$ given by

$$\rho = \rho_0 \left(\frac{R_0}{R}\right)^3$$

and with a mass fraction of $H_2O$ of ~ $10^{-2}$ equation (8) yields

$$\frac{dr}{dt} = \alpha \,(3 \times 10^{-23}) \left(\frac{R_0}{R}\right)^3 \left(\frac{m_{H2O}}{2\pi kT}\right)^{1/2} \tag{9}$$

Setting R=0.1AU, T=100K and $\alpha = 1$ we have

$$\frac{dr}{dt} = 2.4 \times 10^{-4} \; cm/My \tag{10}$$

Grain growth to micron sizes will then take place in < 1 My.

The mass absorption coefficient of such grains for visible light is

$$\kappa = \frac{\pi a^2 Q}{\frac{4}{3}\pi a^3 s} = \frac{3Q}{4as} \cong 10^4 cm^2\, g^{-1}$$

Thus a cloud of diameter 0.2AU has an optical depth in ice grains well in excess of unity. The protoplanetary cloud would then be similar to a Bok globule in our galaxy, with extensive gas phase chemistry taking place within it.

The first stages of prebiotic chemistry would have inevitably begun, with organic molecules sticking onto ice grains to form mantles. Assisted by the reduced gas pressure in the cloud, collisions between organic-coated grains would lead to a rapid collapse on a timescale of < 1My, thus resulting in the formation of a frozen ice-organic planet. For an initial cloud of mass $10^{29}$g with a CNO mass of $10^{27}$g, the radius of the initial ice-organic core with density ~ 1 g cm$^{-3}$ would be ~ 6200km.

## 4.2 Condensation of a solid hydrogen mantle

The condensed icy body will be surrounded by an extended dense halo of molecular hydrogen and helium which would remain largely gaseous until a later epoch, z=6, when the universe cools below the hydrogen triple point 13.8K and the frozen material falls as snow or hail.

The freezing of the $H_2$ envelope as the contraction proceeds much further will depend on the thermodynamic properties of solid $H_2$. The saturation vapour pressure of solid $H_2$ is given to a good approximaton by

$$\log p_{sat} \cong -\frac{40.2}{T} + 2.5 \log T + 6.91 \tag{11}$$

at low temperatures (van de Hulst, 1946; Pfenniger and Puy, 2003). Assuming all the H is as $H_2$, its partial pressure when the contraction is by a factor f is given by

$$p_{H_2}(t) = f^3 \left(3 \times \frac{10^{-21} kT(t)}{m_{H_2}}\right) \qquad (12)$$

taking 3 x $10^{-21}$ g cm$^{-3}$ to be the starting density of the fragment. With T(t) given by equation (5) we can use equations (11) and (12) to calculate the ratio of $p_H/p_{sat}$ as a function of epoch t.

The results of this computation for compressions by factors $10^6$, $3 \times 10^6$, and $5 \times 10^6$ are shown in Fig.5.

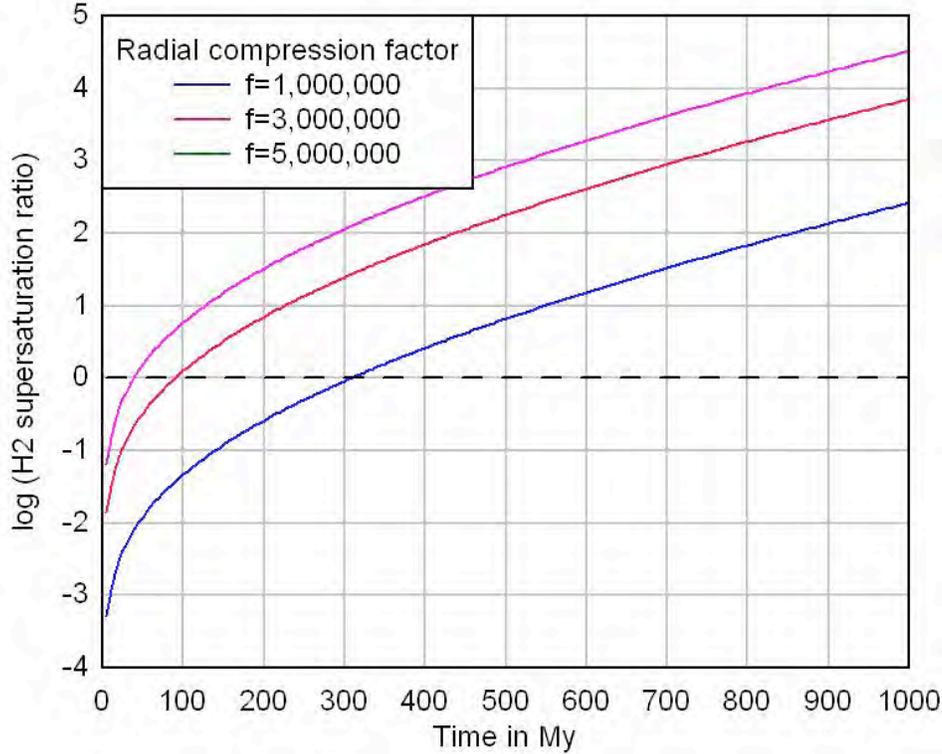

Fig. 5 Supersaturation ratio of molecular hydrogen as a function of epoch for densities enhanced by contraction from the initial value of 3 x $10^{-21}$ g cm$^{-3}$.

A fully-fledged molecular hydrogen planet is not likely to form for ~ $10^2$ My according to this calculation. The outer radius of the planet will be ~ 65,000 km, about 10 times the radius of the frozen water-ice-organic core.

## 5. RADIOACTIVE HEATING AND MELTING OF PLANET CORES

The role of radioactive elements $^{26}Al$ and $^{60}Fe$ in liquefying comet interiors has been discussed in earlier papers (Hoyle and Wickramasinghe, 1982; Wickramasinghe. Wickramasinghe and Wallis, 2009). In solar system comets we estimated that the mass fraction of these elements relative to water-ice was ~ $3 \times 10^{-7}$. In this section we assume that the frozen planet also has properties similar to dirty ice but with a fraction of $^{26}Al$ and $^{60}Fe$ ~ $3 \times 10^{-7}$ f where f is a factor that could be in the range 0.5 – 0.1.

When $H_2O$ ice grains become thermodynamically stable in a contracting cloud at epoch t = 10My, radioactive nuclides generated by supernova explosions would already have been incorporated within the cloud. The radioactive elements $^{26}Al$ and $^{60}Fe$ have half lives typically of 1 My.

Radioactive decays then release heat into the icy cores of the frozen planets. Interior regions of the planets can then become melted, with subsequent cooling and re-freezing taking place when the heat generated is conducted out through overlying ice. Because of the very low thermal conductivity of ice it is possible for large enough icy bodies to maintain liquid interiors for timescales considerably longer than the half-lives of the radionuclides.

A simple model relevant to our problem was discussed by Mark and Prialnik (2003) and Wickramasinghe, Wickramasinghe and Wallis (2009). The heat transfer equation to be solved is

$$C_p \frac{dT}{dt} = \tau^{-1} X_0 H e^{-t/\tau} - \frac{3KT}{R^2 \rho} \tag{13}$$

where T is the average temperature of the body, $C_p$ is the average specific heat, H is the radioactive heat input per unit mass, K is the average thermal conductivity, $X_0$ is the mass fraction of the radioactive elements $^{26}Al$ and $^{60}Fe$ and R is the radius. Here $\tau$ is the average half-life of $^{26}Al$ and $^{60}Fe$ and t is the time. For an ice-organic mix we assume the empirical values/results:

$C_p$ = 7.51T J/kg K
K = 2.22 J/mKs = 600/T J/mKs
T = 1 My for $^{26}Al$ and $^{60}Fe$

Now equation (11) can be solved analytically to yield:

$$T^2 = T_0^2 + \frac{2X_0 H}{a}\left(1 - e^{-\frac{t}{\tau}}\right) - \frac{6bt}{asR^2} \tag{14}$$

Here numerically a=7.51, b=600, $X_0$ = 3x10$^{-7}$ f. The factor f representing a dilution of radionuclides compared with comets in the solar system.

We assume that our planet froze out instantaneously at an epoch $t_1$ somewhere in the range $t_1$ =7-15My. By this time the majority of radioactive nuclides with the shortest half-lives would have long decayed, leaving $^{26}Al$ and $^{60}Fe$ as the main contenders for a dominant heat source in the primordial planet. Since the ensemble of frozen planets in this picture is part of the expanding universe, their outer surface temperature can be assumed to vary with epoch in accordance with equation (5). Setting the value of temperature in equation (5) to be $T_0$, equation (14) can be solved numerically for T(t), the planet's interior temperature, replacing t throughout by t-$t_1$.

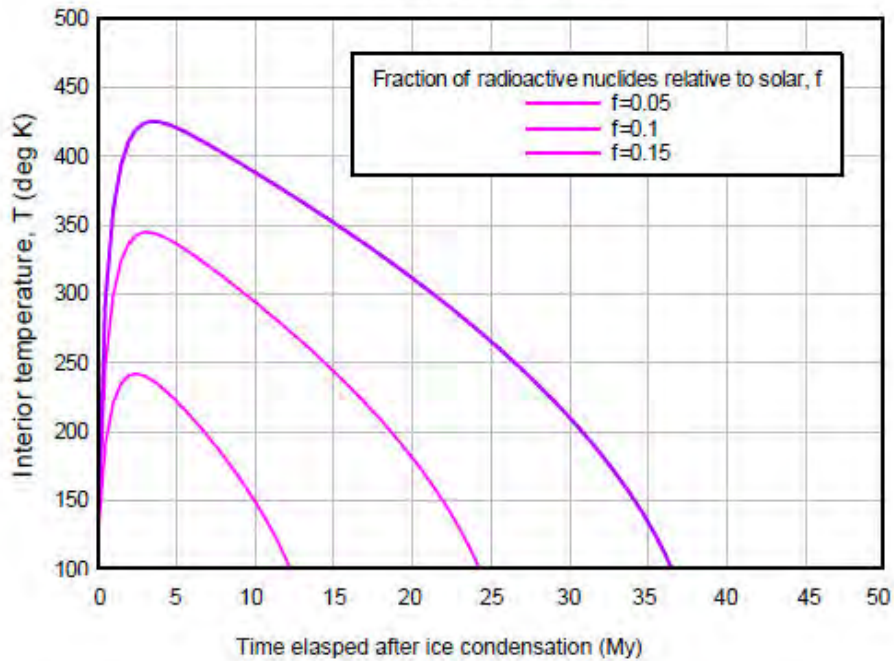

Figure 6a

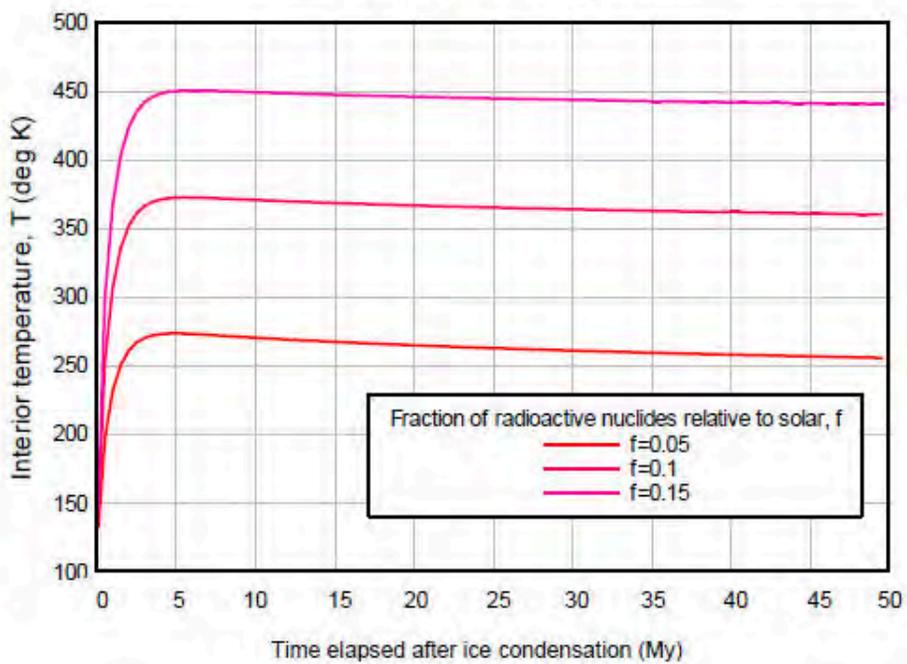

Fig.6b

Figure 6a, b  Interior temperature as a function of epoch for radii of 100 and 1000 km respectively.  The various curves are for different fractions of radioactive nuclides relative to their solar values.

The results of our calculations for R = 100km (planetoid) and 1000km (planet) are plotted in Fig. 6.  In each case we show results for f=0.05, 0.1, 0.15.   For f we consider a value 0.1 to be reasonable, showing here that for the frozen planets (R ≈ 1000km) of the type discussed by Gibson and Schild we have interior temperatures maintained well above the triple point of water for periods of time exceeding 10My.  Each such planet would then serve as a gigantic primordial soup well suited for abiogenesis in the manner proposed by Oparin and Haldane.

Moreover, with average distances between such planets being estimated to be a few AU at the epoch t=10My, frequent exchanges of organic material and evolving life-templates would enlarge the scale of the soup to encompass essentially the entire universe.  No other setting at any later epoch could remotely match the advantages offered at this unique and crucial moment in the history of the cosmos.

## 6.  PROPOSED SCHEMATIC TIMETABLE FOR CHEMICAL EVOLUTION

Following the onset of instability in the HGD model of Gibson and Schild, the subsequent condensation and chemical history of a typical planetary mass cloud is summarised in Table I.  Here we trace the history of a cloud 50% higher than the ambient density at the plasma-neutral transition -  i.e. with a density of $3 \times 10^{-21}$ g cm-3, and an initial radius of ~ 300AU at the time of their formation, 0.3My after the Big Bang.

Table 1

Gibson-Schild instability sets in at t=0.3My

| Epoch t (My) | Redshift (z) | Background Temperature, (degK) | Ambient density of Universe ($g\ cm^{-3}$) | Radius of protoplanetary cloud/planet | Comment |
|---|---|---|---|---|---|
| 0.3 | ~1200 | 2230 | $2 \times 10^{-21}$ | 300AU | Collapse begins |
| 0.3-2 | 1200-300 | 2230-647 | $2 - 0.19 \times 10^{-21}$ | <300AU | Runaway accretion leading to SN; condensation of mineral grains, first PFP life |
| 7-15 | 180-100 | 312-187 | $1 - 4 \times 10^{24}$ | 3AU – 0.3AU | Water-ice thermodynamically stable, ice grains form; icy core forms |
| 10-60 | 140-43 | 245-74 | $190 - 5.21 \times 10^{-26}$ | Planet radius 6,200km; | Radioactive heat keeps much of interior melted; primordial soup for the origin of life on solar nebula planets |
| 100 | ~30 | 52 | $1.9 \times 10^{-26}$ | Planet radius grows rapidly above 62,000km | Solid hydrogen is thermodynamically stable and forms solid grains |
| 1000 | ~6 | 13.8 | $1.9 \times 10^{-28}$ | Final planet radius 65,000km | Solid hydrogen falls as hail/snow increasing radius by ~10 |

## 7.  CONCLUDING REMARKS

An important future task of astrobiology is to model the conditions within primordial planets most likely to lead to the first life and its cosmic spreading by cometary panspermia.  It is also important to explore faster routes to the emergence of life within the collapsing fragments if such routes are feasible.  The earlier the origin of life occurs, the more likely is its initial spread across the cosmos, because of the initial high density and rapid expansion of the Universe.  A proper fluid mechanical modelling of the collapse including effects of turbulence driven by radiative cooling at the surface, and convection, may lead to the production of a high pressure dense core replete with liquid water and organic molecules.  This may be possible to form, under suitable condition, within the gravitational collapse timescale of ~ 1 My.

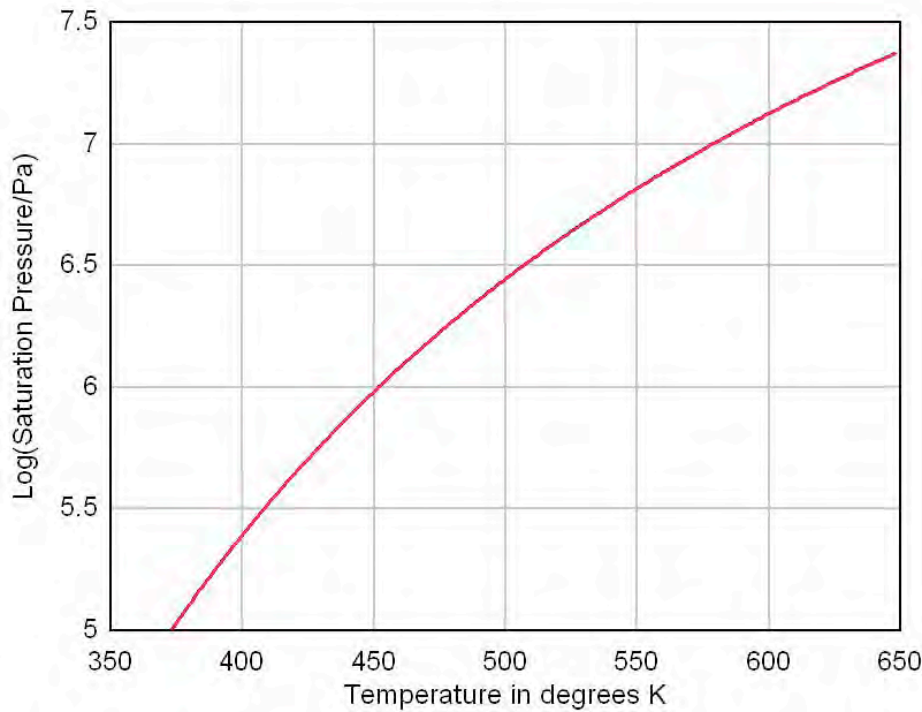

Fig. 7 Saturation vapour pressure of liquid water from the triple point to the critical temperature of 650K

Figure 7 shows the saturation vapour pressure of liquid water for the temperature range from 273 – 650K, ie from the triple point to the critical point. From Fig.7 we note that the value of saturation vapour pressure at 650K (i.e. at epoch t=2My according to Fig. 2) is $10^{18}$ times the water vapour pressure at t=0.3My when the cloud fragment begins its collapse. If such a high pressure central core is feasible then it will be possible to contemplate an early formation of liquid water domains, and an earlier origin of life. Density enhancements towards the centre of a polytrope of index n=∞ could be as high as $10^{18}$, and such a model might possibly apply to the case of an isothermal gas sphere in hydrostatic equilibrium (Eddington, 1926). Water and organics would then remain close to the critical temperature for millions of years due to the insulating properties of the outer crust.

In conclusion we note that a planetary mass object that presumably recently formed or was modified, but which would have properties similar to our primordial planet population, may have been observed by Marsh, Kirkpatrick and Plavchan (2010). Their 2-3MJup object's measured surface temperature 1400K and the inferred evolutionary age ~ 1My correspond to primordial planets in the 0.3-2My epoch ($2^{nd}$ row of Table 1). This discovery can arguably be taken as evidence for freezing of ices earlier than later as proposed in our 2-10My window.